\documentstyle[amssymb,epsfig]{aipproc}

\begin{document}

\title{Broadband dielectric response of glycerol and propylene carbonate: a
comparison }
\author{P. Lunkenheimer, U. Schneider, R. Brand, and A. Loidl}
\address{Experimentalphysik V, Universit\"{a}t Augsburg, D-86135 Augsburg,
Germany}
\maketitle

\begin{abstract}
Dielectric data on glycerol and propylene carbonate covering 18 decades of
frequency are presented and compared to each other. Both materials exhibit
qualitatively similar behavior except for marked differences in the
high-frequency region just below the boson peak. The results on both
materials are consistent with the mode coupling theory of the glass
transition.
\end{abstract}

\section{Introduction}

In recent years, a variety of new theoretical and phenomenological
approaches of the glass transition (e.g., \cite{mctrev,NgaiKiv,Nagscal})
stimulated new experimental investigations especially of the high-frequency
dynamics of glass-forming liquids \cite
{Men,nsrev,Cum,Wutt,Du,Le,Lunkigly,Lunkiorl,Lunkikyockn,Lunkibos,Schn,Schnpc}%
. Maybe the most controversially discussed theoretical approach of the glass
transition is the mode coupling theory (MCT) \cite{mctrev}, which
explains the glass transition in terms of a dynamic phase transition
at a critical temperature $T_c$significantly above the glass
temperature $T_g$. For frequencies in the GHz-THz region, MCT predicts
an additional contribution, now commonly termed fast $\beta
$-relaxation. This frequency region was mainly investigated by neutron
and light scattering experiments \cite{nsrev,Cum,Wutt,Du}. But
recently, by combining various techniques, our group was able to
obtain continuous dielectric spectra on glass-forming liquids
extending well into the relevant region \cite
{Lunkigly,Lunkiorl,Lunkikyockn,Lunkibos,Schn,Schnpc}. For
glass-forming glycerol and propylene carbonate, spectra covering 18
decades of frequency and extending well into the THz range were
obtained \cite{Schn,Schnpc}. Glycerol is a rather strong \cite{strong}
hydrogen-bonded glass-former with a fragility parameter of $m\approx
53$ \cite{rol}. In contrast, propylene
carbonate (PC) can be characterized as a fragile ($m\approx 104$, \cite{rol}%
) van der Waals liquid. In the present paper we present a comparison of the
results on both materials. For the experimental details the reader is
referred to our earlier publications \cite{Lunkiorl,Schn,Schnpc}.

\section{\protect\bigskip Results and discussion}

\begin{figure}[Ht]
\centerline{\epsfig{file=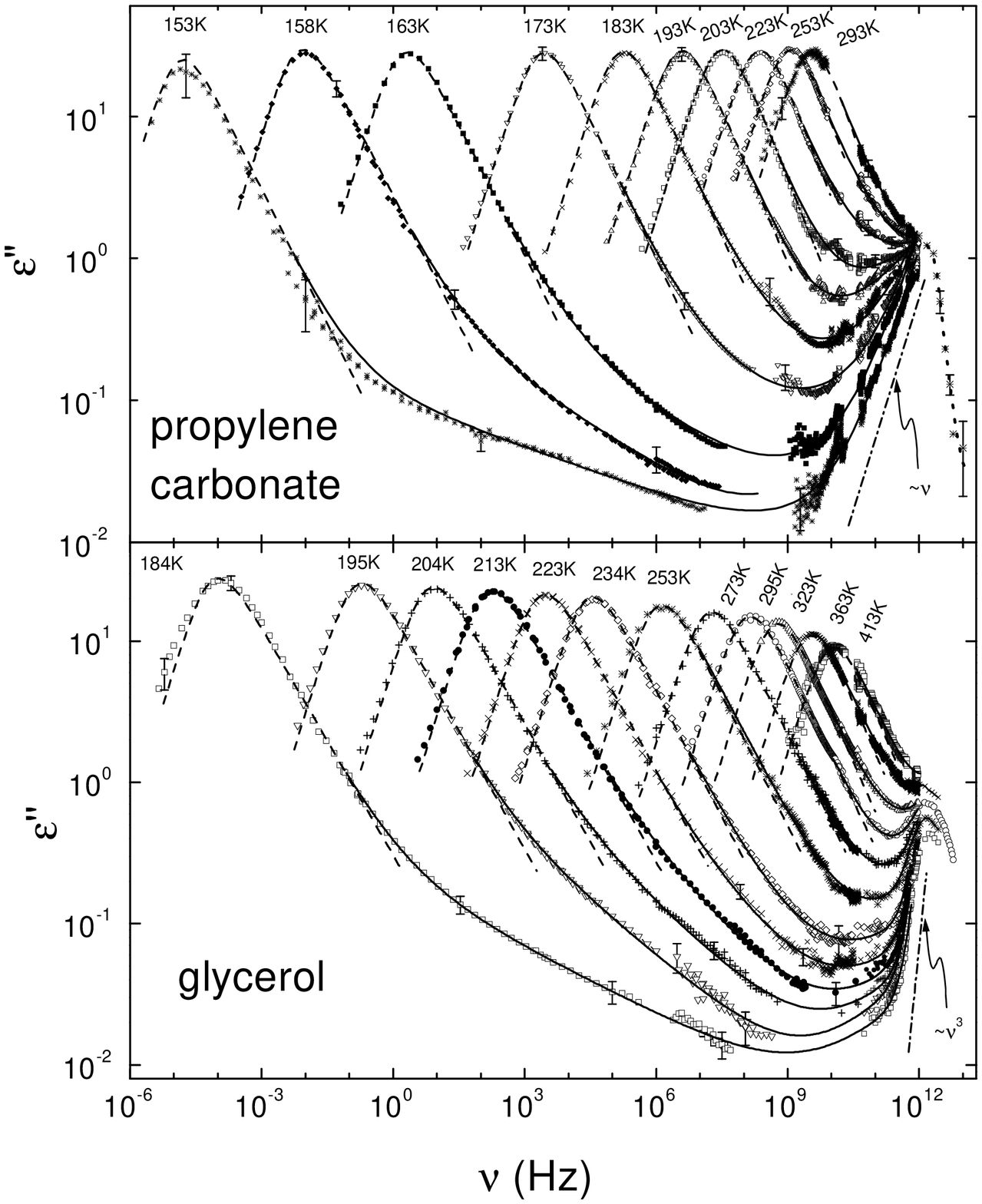,width=10cm}} \vspace{10pt}
\caption{$\protect\varepsilon ^{\prime \prime }(\protect\nu )$ of PC (a) and
glycerol (b) at various temperatures. The dashed lines are fits of the $%
\protect\alpha $-peak with the CD function. The solid lines are fits with a
phenomenological ansatz (see text). The dash-dotted lines indicate a power
law increase at $\protect\nu \lesssim 1$ THz. The FIR results in (a) have
been connected by a dotted line to guide the eye. }
\label{figeps2}
\end{figure}

\begin{figure}[Ht]
\centerline{\epsfig{file=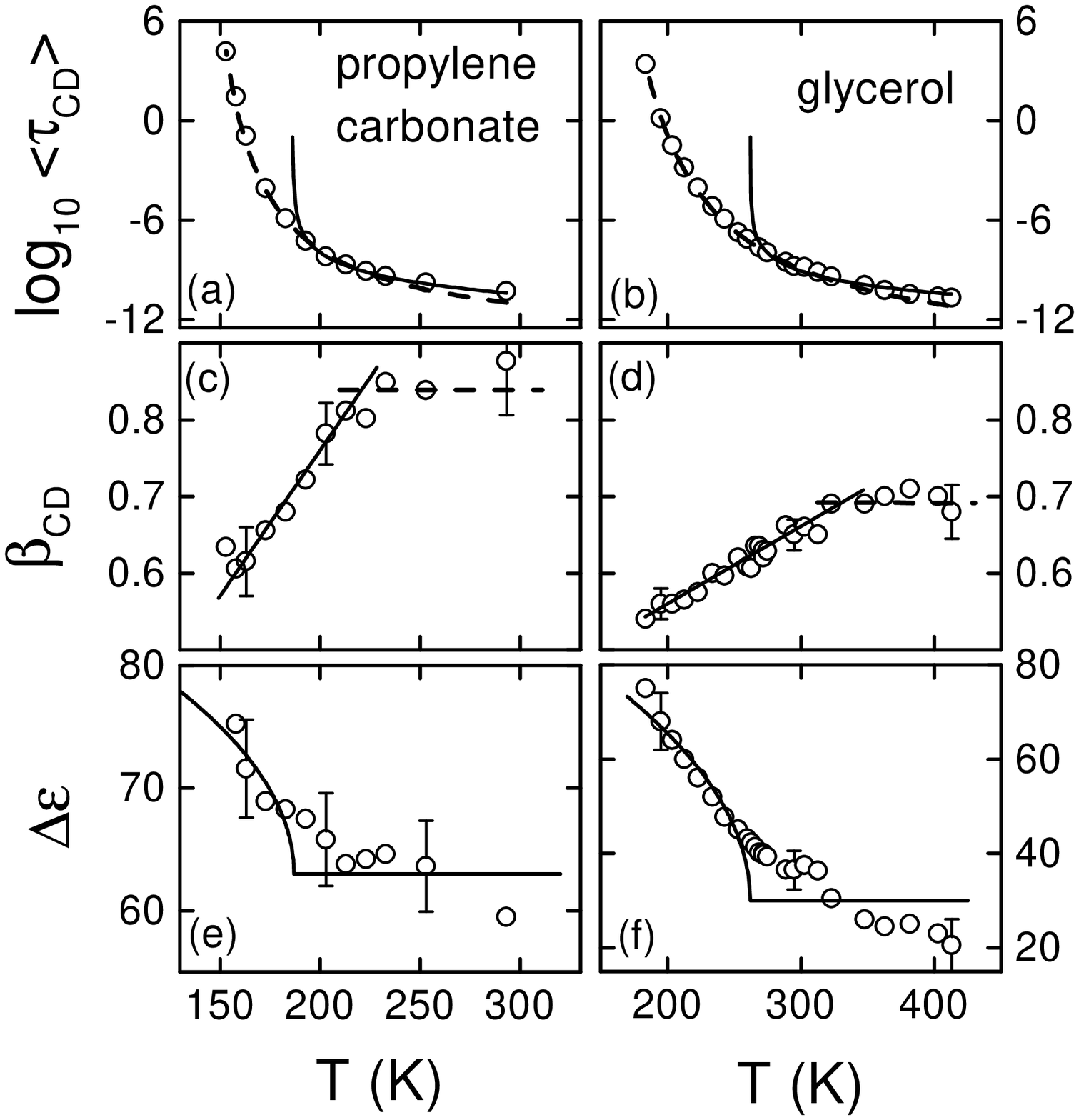,width=9cm,width=3.5in}} \vspace{10pt}
\caption{Parameters from simultaneous fits of $\protect\varepsilon ^{\prime
}(\protect\nu )$ and $\protect\varepsilon ^{\prime \prime }(\protect\nu )$\
with the CD function. The dashed and solid lines in (a) and (b) are fits
with the VFT law and the MCT critical behavior, respectively. The lines in
(b) and (c) demonstrate a possible saturation of $\protect\beta _{CD}(T)$ at
high temperatures. The lines in (e) and (f) are calculated according to MCT
(see text).}
\label{figcdfipa}
\end{figure}

Figure \ref{figeps2}
shows the $\varepsilon ^{\prime \prime }(\nu )$-spectra for glycerol
and PC in the whole accessible frequency range. $\varepsilon ^{\prime
\prime }(\nu ) $ exhibits the typical asymmetrically shaped $\alpha
$-relaxation peaks shifting through the frequency window with
temperature. The data agree with the results of earlier dielectric
investigations \cite {Men,Le,dielPC,AngPC,CD} which were restricted to
smaller frequency and
temperature ranges. The dashed lines in Fig. \ref{figeps2} are fits of the $%
\alpha $-relaxation region with the empirical Cole-Davidson function \cite
{CD}. A good fit of the peak region was achieved. The temperature dependence
of the parameters of the CD function, $<\tau _{CD}>=\tau _{CD}\beta _{CD}$, $%
\beta _{CD}$, and $\Delta \varepsilon =$ $\varepsilon _{s}-\varepsilon
_{\infty }$ \cite{CD} is shown in Fig. \ref{figcdfipa}.
In most respects, the results are in accord with previously published
data \cite{Men,dielPC,AngPC,CD,Stick2} extending them to lower or
higher temperatures. $\nu _{\tau }(T)$ can be parameterized using the
Vogel-Fulcher-Tamman (VFT) equation, $<\tau _{CD}>=\tau _{0}\exp
[DT_{VF}/(T-T_{V\!F})]$ with $T_{V\!F}=$ $132$ K, $D=6.6$ for PC and $%
T_{V\!F}=131$ K, $D=16.5$ for glycerol [dashed lines in Figs. \ref{figcdfipa}%
(a) and (b)]. The values of the strength parameter $D$ characterize PC
as fragile and glycerol as intermediate in Angell's classification
scheme \cite{strong}. At high temperatures small deviations from VFT
behavior show up, similar to those seen in earlier work \cite
{Men,AngPC,Stick2}. The simplest version of MCT, the idealized MCT
\cite {mctrev}, predicts a critical behavior of the $\alpha
$-relaxation timescale, $\nu _{\tau }=1/(2\pi <\tau _{CD}>)\thicksim
(T-T_{c})^{\gamma }$ with $\gamma =1/(2a)+1/(2b)$. Here $a$ and $b$
are the power law exponents of the $\varepsilon ^{\prime \prime }(\nu
)$-minimum (see below). The solid lines in Figs. \ref{figcdfipa}(a)
and (b) are fits with the MCT prediction with $\gamma =2.72,$
$T_{c}=187$ K for PC and $\gamma =2.33,$ $T_{c}=262$ K
for glycerol, fixed to the values obtained from the analysis of the $%
\varepsilon ^{\prime \prime }(\nu )$-minimum (see below). Only the data
above 200 K for PC and above 280 K for glycerol have been used for the fits
which led to a good agreement in this temperature region. For lower
temperatures deviations show up. This can be understood within the extended
MCT \cite{mctrev} where the structural arrest found in idealized MCT for $%
T<T_{c\text{ }}$is overcome by the introduction of thermally activated
hopping processes.

For both materials, the width parameter $\beta _{CD}(T)$ [Figs. \ref
{figcdfipa}(c) and (d)] increases linearly below about 200 K for PC and 300
K for glycerol. For $T>T_{c}$, MCT predicts a temperature independent
spectral form of the $\alpha $-process. Indeed, a tendency to saturate is
seen for $\beta _{CD}(T)$ in Figs. \ref{figcdfipa}(c) and (d), however at
temperatures somewhat above $T_{c}$, only. But a reasonable description of
the $\alpha $-peaks for $T>T_{c}$ is also possible with a constant $\beta
_{CD}=0.63$ for glycerol \cite{Lunkibos} and $\beta _{CD}=0.80$ for PC. The $%
\Delta \varepsilon (T)$ of PC and glycerol [Figs. \ref{figcdfipa}(e) and
(f)] both decrease with temperature. According to MCT, $\Delta
\varepsilon $ should be temperature independent for $T>T_{c}$. In
addition, for $T<T_{c}$, $\Delta \varepsilon
=c_{1}+c_{2}[(T_{c}-T)/T_{c}]^{\frac12}$ follows from extended MCT. In Figs. \ref{figcdfipa}(e) and (f) the lines
were calculated using the MCT prediction. The agreement of fits and data is\
rather poor, but due to the large experimental uncertainties no definite
conclusion can be drawn. At least it seems that $\Delta \varepsilon $
assumes a weaker temperature dependence above $T_{c}$.

At frequencies about 2-3 decades above the peak frequency, for both
materials an excess wing shows up as power law, $\varepsilon ^{\prime \prime
}\thicksim \nu ^{-b}$ with $b<\beta _{CD}$ (Fig. \ref{figeps2}). The
exponent $b$ decreases with decreasing temperature as found previously for
various glass-forming materials \cite{Le}. For PC, at 203 K and for glycerol
at 272 K, the excess wing has merged with the $\alpha $-peak. The excess
wing seems to be a universal feature of glass-forming liquids \cite{Nagscal}
but, up to now, no consensus concerning its microscopic origin was achieved.
For a more detailed treatment of the excess wing in glycerol and PC, see
\cite{Schnpc}.

\begin{figure}[Ht]
\centerline{\epsfig{file=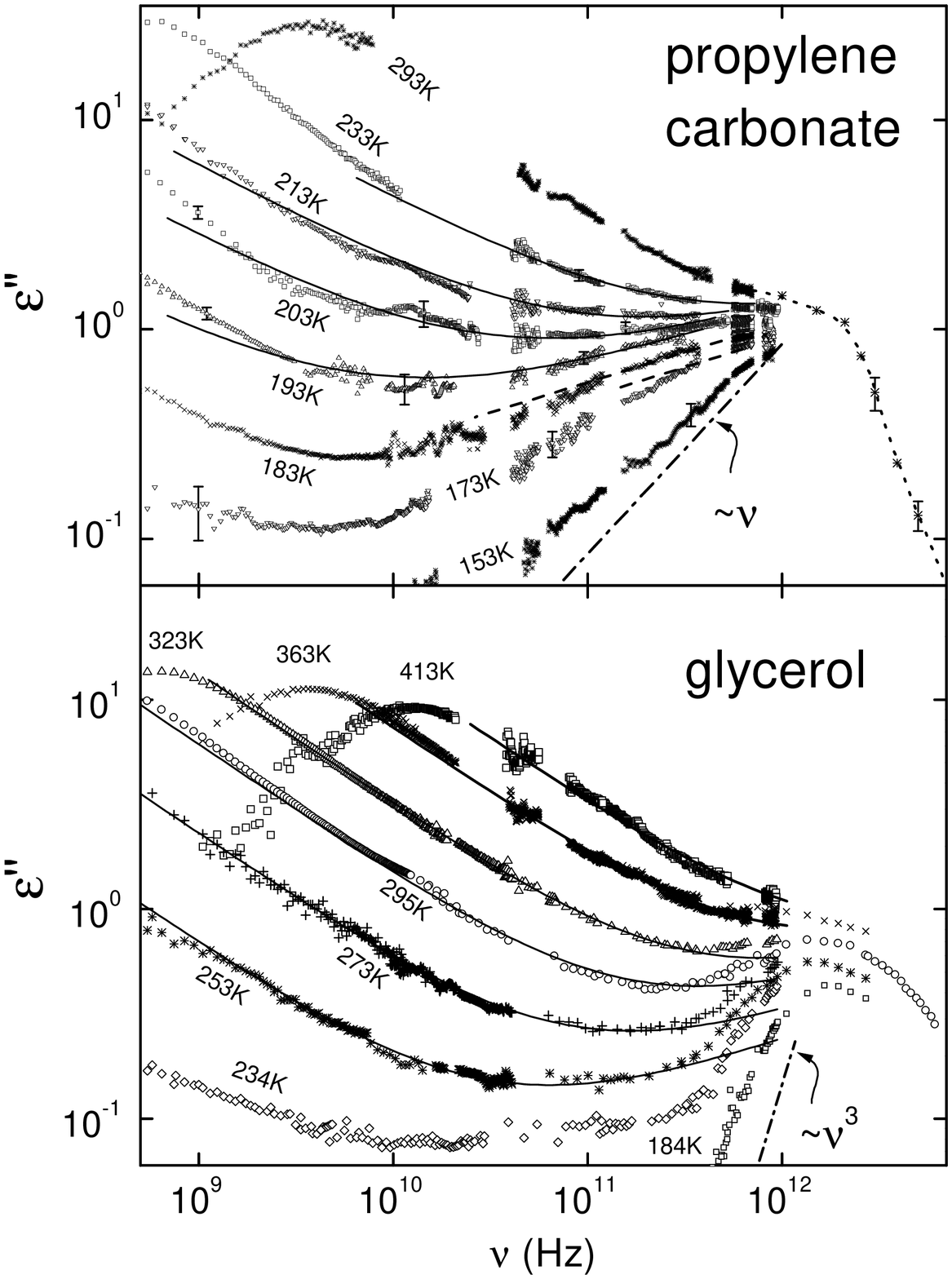,width=9cm}} \vspace{10pt}
\caption{$\protect\varepsilon ^{\prime \prime }(\protect\nu )$ of PC (a) and
glycerol (b) at high frequencies. The solid lines are fits with the MCT
prediction (see text). The dashed lines demonstrate $\protect\nu ^{a}$
behavior for $T<T_{c}$. The dash-dotted lines indicate a power law increase
at $\protect\nu \lesssim 1$ THz. The FIR results in (a) have been connected
by a dotted line to guide the eye. }
\label{figmct}
\end{figure}

At frequencies in the GHz-THz range, $\varepsilon ^{\prime \prime }(\nu )$
exhibits a minimum for both materials \cite{Lunkigly,Lunkiorl}. A magnified
view is given in Fig. \ref{figmct}.
Clearly this minimum is too shallow to be described by a simple
superposition of $\alpha $-peak (or excess wing) and boson peak \cite
{Lunkigly,Lunkiorl}. Near 1 THz, for both materials the FIR measurements
reveal a peak or shoulder, that can be ascribed to the boson peak, well
known from light and neutron scattering experiments \cite{Wutt,Du}. Both
materials exhibit a significantly different behavior in the region between
minimum and boson peak: For PC only one power law, $\varepsilon ^{\prime
\prime }\thicksim \nu ^{a}$, is seen at frequencies $\nu >\nu _{\min }$
forming simultaneously the high frequency wing of the $\varepsilon ^{\prime
\prime }$-minimum and the low frequency wing of the boson peak. For $%
T\rightarrow T_{g}$, a linear behavior is approached as indicated by the
dash-dotted lines in Figs. \ref{figeps2}(a) and \ref{figmct}(a). In marked
contrast, for glycerol two regimes can be distinguished: Just above $\nu
_{\min }$, there is a rather shallow increase of $\varepsilon ^{\prime
\prime }(\nu )$\ but at higher frequencies a very steep increase appears,
approaching $\varepsilon ^{\prime \prime }\thicksim \nu ^{3}$ for $%
T\rightarrow T_{g}$ [dash-dotted lines in Figs. \ref{figeps2}(b) and \ref
{figmct}(b)].

For both materials, the $\varepsilon ^{\prime \prime }(\nu )$-spectra
between $\alpha $- and boson peak can be described by the phenomenological
ansatz $c_{\beta }\nu ^{-\beta }+c_{b}\nu ^{-b}+\varepsilon _{c}+c_{3}\nu
^{0.3}+c_{n}\nu ^{n}$ (solid lines in Fig. \ref{figeps2}). The first term
takes account of the high-frequency wing of the $\alpha $-peak with the
parameters chosen to achieve a smooth transition to the CD-fits. The second
term describes the excess wing. In order to explain the shallow minimum, a
constant loss term, $\varepsilon _{c}$, as proposed by Wong and Angel \cite
{Wong} and a $\varepsilon ^{\prime \prime }\thicksim \nu ^{0.3}$ increase
were assumed. Both contributions seem to be universal features in the
high-frequency response of glassy ionic conductors \cite{colo03}. Finally,
the $\nu ^{n}$ power law, with $n\geq 1,$ takes account of the low frequency
wing of the boson peak. In this way, quite impressive fits over 17 decades
of frequency are possible. Of course this could be expected considering the
large amount of parameters, but nevertheless it cannot be fully excluded
that such a simple superposition ansatz is correct.

MCT predicts for the minimum region significant contributions from the fast $%
\beta $-process. Within idealized MCT, above $T_{c}$, the minimum region can
be approximated by the interpolation formula \cite{mctrev} $\varepsilon
^{\prime \prime }(\nu )=\varepsilon _{\min }[a(\nu /\nu _{\min })^{-b}+b(\nu
/\nu _{\min })^{a}]/(a+b)$. Here $\nu _{min}$ and $\varepsilon
_{min}^{\prime \prime }$ denote frequency and amplitude of the minimum,
respectively. The exponents $a$ and $b$ are temperature independent and
constrained by $\lambda =\Gamma ^{2}(1-a)/\Gamma (1-2a)=\Gamma
^{2}(1+b)/\Gamma (1+2b)$ where $\Gamma $ denotes the Gamma function. This
implies $a<0.4$, i.e. a significantly sublinear increase of $\varepsilon
^{\prime \prime }(\nu )$ at $\nu >\nu _{min}$ is predicted. For PC, a
consistent description of the $\varepsilon ^{\prime \prime }$-minima at $%
T\geq 193$ K is possible with $\lambda =0.76$ ($a=0.29$, $b=0.5)$ \cite
{Lunkiorl,Schnpc} [solid lines in Fig. \ref{figmct}(a)]. The obtained $%
\lambda $ is consistent with the results from other measurement techniques
\cite{Du,Ohl,Berg}. The fits provide a good description of the data up to
the boson-peak frequency. Quite a different behavior is seen in glycerol
[solid lines in Fig. \ref{figmct}(b)]: Here the MCT fits ($\lambda =0.705$, $%
a=0.325$, $b=0.63)$ \cite{Lunkigly,Lunkiorl} are limited at high frequencies
by the additional steeper increase, mentioned above. It was argued \cite
{Lunkigly,Lunkiorl}, that this deviation in glycerol is due to the
vibrational boson peak contribution which is not taken into account by the
MCT interpolation formula. In PC this contribution seems to be of less
importance. This is in accord with the finding of Sokolov {\it et al}. \cite
{Sok} that the amplitude ratio of boson peak and fast process is largest for
strong glass formers, glycerol being much stronger than PC. However, it has
to be noted, that in the present results, at least near $T_{g}$ this ratio
is quite similar for both materials.

\begin{figure}[Ht]
\centerline{\epsfig{file=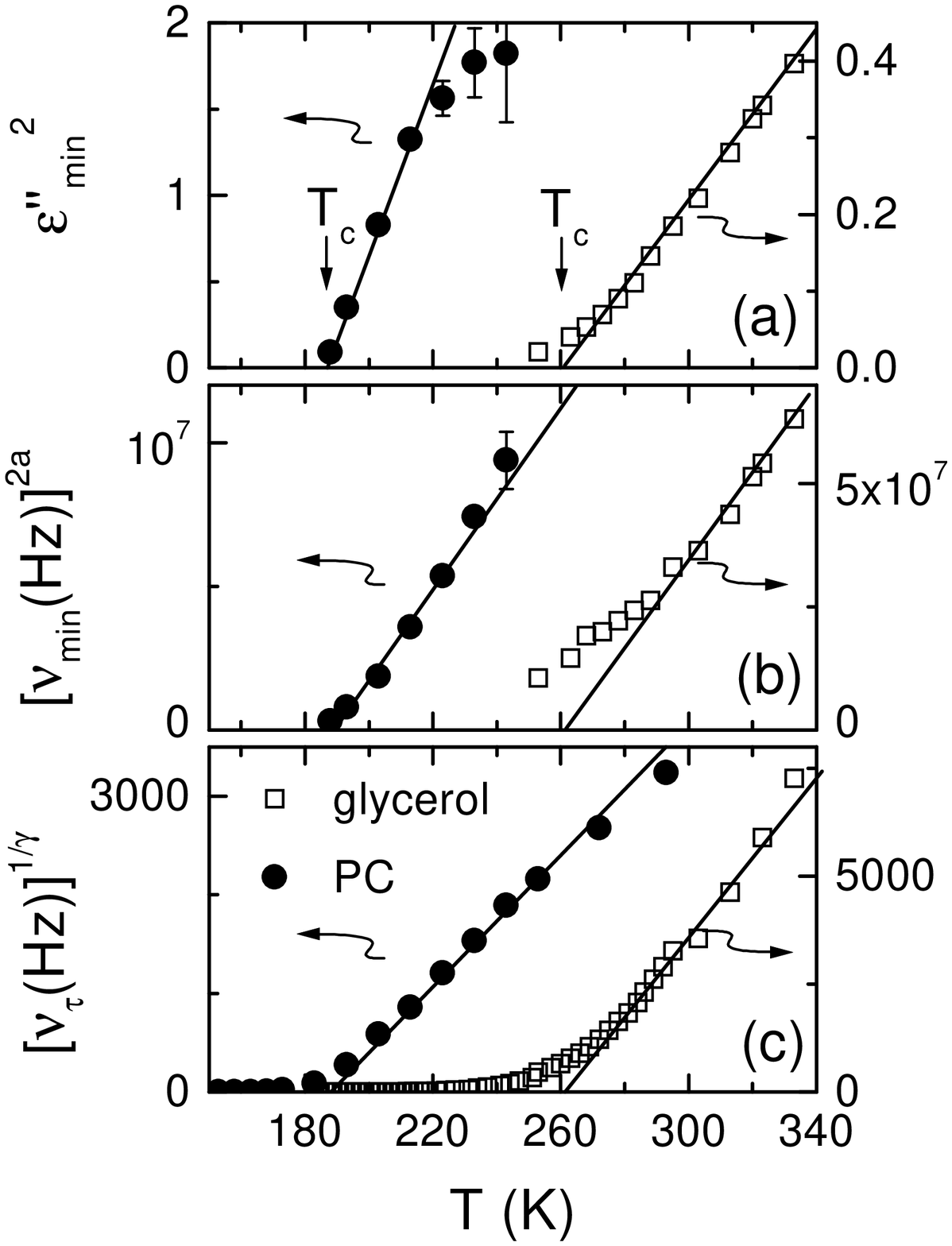,width=9cm}} \vspace{10pt}
\caption{Temperature dependence of $\protect\varepsilon _{\min }^{\prime
\prime }$, $\protect\nu _{\min }$, and $\protect\nu _{\protect\tau }$
plotted in representations that result in straight lines according to the
predictions of MCT. The solid lines extrapolate to a $T_{c}$ of 187 K for PC
and 262 K for glycerol.}
\label{figmopa}
\end{figure}

For $T>T_{c}$, idealized MCT predicts the following critical temperature
dependences: $\nu _{min}\thicksim (T-T_{c})^{1/(2a)}$, $\varepsilon
_{min}^{\prime \prime }\thicksim (T-T_{c})^{\frac12}$, and $\nu _{\tau }\thicksim (T-T_{c})^{\gamma }$. In Figure \ref{figmopa}
representations have been chosen that should lead to straight lines
that extrapolate to $T_{c}$. Indeed, all three data sets can be
described
consistently with a $T_{c}\approx 187$ K for PC \cite{Lunkiorl,Schnpc}\ and $%
T_{c}\approx 262$ K for glycerol \cite{Lunkigly} as indicated by the solid
lines. These values lie in the same range as the $T_{c}$-value obtained from
other techniques \cite{Wutt,Du,Ohl,Berg,Borj}. For temperatures near $T_{c}$
the data partly deviate from the predicted behavior. Within MCT this can be
ascribed to a smearing out of the critical behavior near $T_{c}$ due to
hopping processes which are considered in extended versions of MCT only \cite
{mctrev}. In addition, the above critical temperature dependences of $%
\varepsilon _{min}^{\prime \prime }$, $\nu _{min}$, and $\nu _{\tau }$
should be valid only for temperatures not too far above $T_{c}$. Therefore
the proper choice of the temperature range to be used for the determination
of $T_{c}$ is difficult which may lead to some uncertainties concerning the
value of $T_{c}$.

For $T<T_{c}$, MCT predicts the occurrence of a so-called ''knee'' at $\nu =$
$\nu _{k}$\ showing up as a change of power law from $\varepsilon ^{\prime
\prime }\sim \nu ^{a}$ at $\nu >$ $\nu _{k}$ to $\varepsilon ^{\prime \prime
}\sim \nu $ at $\nu <\nu _{k}$ \cite{mctrev}. In Fig. \ref{figmct}(a) the
dashed lines suggest a $\varepsilon ^{\prime \prime }\thicksim \nu ^{a}$
behavior for frequencies just below the boson peak. For lower frequencies
the experimental data exhibit a downward deviation from the $\nu ^{a}$%
-lines. This may be indicative of the ''knee'' predicted by MCT, but,
clearly, the data are not of sufficient quality to demonstrate conclusively
the existence of the ''knee''.

\section{Conclusions}

Dielectric data on glass-forming PC and glycerol in an exceptionally broad
frequency range have been presented and compared to each other and with the
predictions of MCT. Concerning the $\alpha $-relaxation and the excess wing,
both materials behave qualitatively similar. The parameters describing the
spectral form and dynamics of the $\alpha $-relaxation are in rough
agreement with the MCT predictions but clearly, Fig. \ref{figcdfipa} is not
very convincing in this respect. The high-frequency response of both
materials provides clear evidence for the presence of fast processes in the
GHz-THz region. For glycerol an additional very steep increase towards the
boson peak seems to be superimposed to the shallow $\varepsilon ^{\prime
\prime }(\nu )$-minimum. For the more fragile PC a smooth sublinear increase
of $\varepsilon ^{\prime \prime }(\nu )$\ is seen up to the boson peak
frequency. For both materials the frequency and temperature dependence in
the minimum region can be consistently described by MCT. Admittedly, also
alternative explanations may be possible. But even in its idealized form,
MCT provides a consistent picture for a large variety of experimental facts.
Obviously, more sophisticated approaches within MCT, e.g. including hopping
processes \cite{mctrev}, the boson peak \cite{Fra}, or orientational degrees
of freedom \cite{Schkug} are necessary to eliminate deviations still seen in
the present work.

\acknowledgments

We gratefully acknowledge stimulating discussion with C.A. Angell, R.
B\"{o}hmer, H.Z. Cummins, W. G\"{o}tze, K.L. Ngai, R. Schilling, W.
Schirmacher, and J. Wuttke. This work was supported by the Deutsche
Forschungsgemeinschaft, Grant-No. LO264/8-1 and the BMBF, contract-No.
13N6917.

\end{document}